\documentstyle[aps,prl,multicol,epsfig]{revtex}
\begin{document}
\draft
\title{Quantum mechanical relaxation of open quasiperiodic systems}
\author{A. Ossipov, M. Weiss, Tsampikos Kottos, and T. Geisel
\\
Max-Planck-Institut f\"ur Str\"omungsforschung und Fakult\"at Physik
der Universit\"at G\"ottingen,\\
Bunsenstra\ss e 10, D-37073 G\"ottingen, Germany}
\maketitle

\begin{abstract}
We study the time evolution of the survival probability $P(t)$ in open 
one-dimensional quasiperiodic tight-binding samples of size $L$, at critical
conditions. We show that it decays algebraically as $P(t)\sim t^{-\alpha}$
up to times $t^*\sim L^{\gamma}$, where $\alpha = 1-D_0^E$, $\gamma=1/D_0^E$
and $D_0^E$ is the fractal dimension of the spectrum of the closed system. 
We verified these results for the Harper model at the metal-insulator 
transition and for Fibonacci lattices. Our predictions should be observable 
in propagation experiments with electrons or classical waves in quasiperiodic 
superlattices or dielectric multilayers.
\end{abstract}

\begin{multicols}{2}

The decay properties of open quantum mechanical systems, have been attracting 
considerable attention for several decades. Their study was motivated by various 
areas of physics, ranging from nuclear \cite{nuclear}, atomic \cite{atomic} 
and molecular \cite{molecular} physics, to mesoscopics \cite {mesoscopics} and 
classical wave scattering \cite{KS00}. In recent years, the interest in quantum 
mechanical decay was stirred by mesoscopic cavities and microwave billiards 
where immediate experimental realizations have become feasible \cite{GSST99}. 
At the same time various analytical techniques have been developed to study 
the problem in more detail. One possible formulation of the problem is to 
consider the survival probability $P(t)$ of a wave packet localized initially 
inside the open system. In particular it was found that this quantity exhibits
slower than exponential decay (i.e. long-time tails) for disordered wires in 
the localized (in one-dimension) \cite{AKL87} and metallic (in higher dimensions) 
regimes \cite{MK95,M95,FE95,SA97}. Moreover, for ballistic systems, Random 
Matrix Theory (RMT) predicts an algebraic decay $P(t) \sim 1/ t^{\beta M/2}$, 
where $M$ is the number of channels and $\beta=1(2)$ for preserved (broken) 
time reversal symmetry \cite{SS97,D00}.

The investigation of the survival probability has recently been extended to quantum 
systems with a mixed classical phase space \cite{CMS99,HKW00} where it was found
that $P(t)\sim1/t$. The same algebraic decay was found for systems with exponential 
localization. In both cases, this law is related to localization and tunneling 
effects and applies for intermediate asymptotic times \cite{CMS99}.

The subject of the present paper is the survival probability
in a new setting, namely a class of systems whose closed analogues have
fractal spectra. The latter exhibit level clustering \cite{GKP95} in strong
contrast to the level repulsion predicted by RMT for systems in the
ballistic regime and to the Poisson statistics in the localized
regime \cite{P65}. Representatives of this
class are quasi-periodic systems with a metal-insulator transition
like the Harper model \cite{GKP95,AA80},
Fibonacci chains \cite{GKP95,SBGC84}, or quantum systems with a 
chaotic classical limit like the kicked Harper model \cite{GKP91}.

Here for the first time we present consequences of the fractal nature of the 
spectrum for the quantum time evolution of open systems. In particular, we 
ask how they are encoded in the survival probability $P(t)$, which is the
simplest quantity measured in laboratory experiments. We show that $P(t)$ 
decays as
\begin{eqnarray}
\label{powlaw}
P(t) \sim 1/t^{\alpha}\,;\,\,\,\;\;\alpha=1-D_0^E,
\end{eqnarray}
where $D_0^E$  is the fractal (box-counting) dimension of the spectrum of the
closed system. Moreover, we determine the time scale $t^*$ up to which this
power-law decay can be observed. It scales as 
\begin{equation}
\label{maxt}
t^*  \sim L^{\gamma}\,;\,\,\,\;\; \gamma=1/D_0^E,
\end{equation}
where $L$ is the sample size. Beyond this time scale $P(t)$ decays 
exponentially. Our results (\ref{powlaw}),(\ref{maxt}) 
are confirmed numerically for two types of quasi-periodic 
tight-binding models and are  supported by analytical
arguments. 

\begin{figure}
\hspace*{-0.7cm}
\epsfig{figure=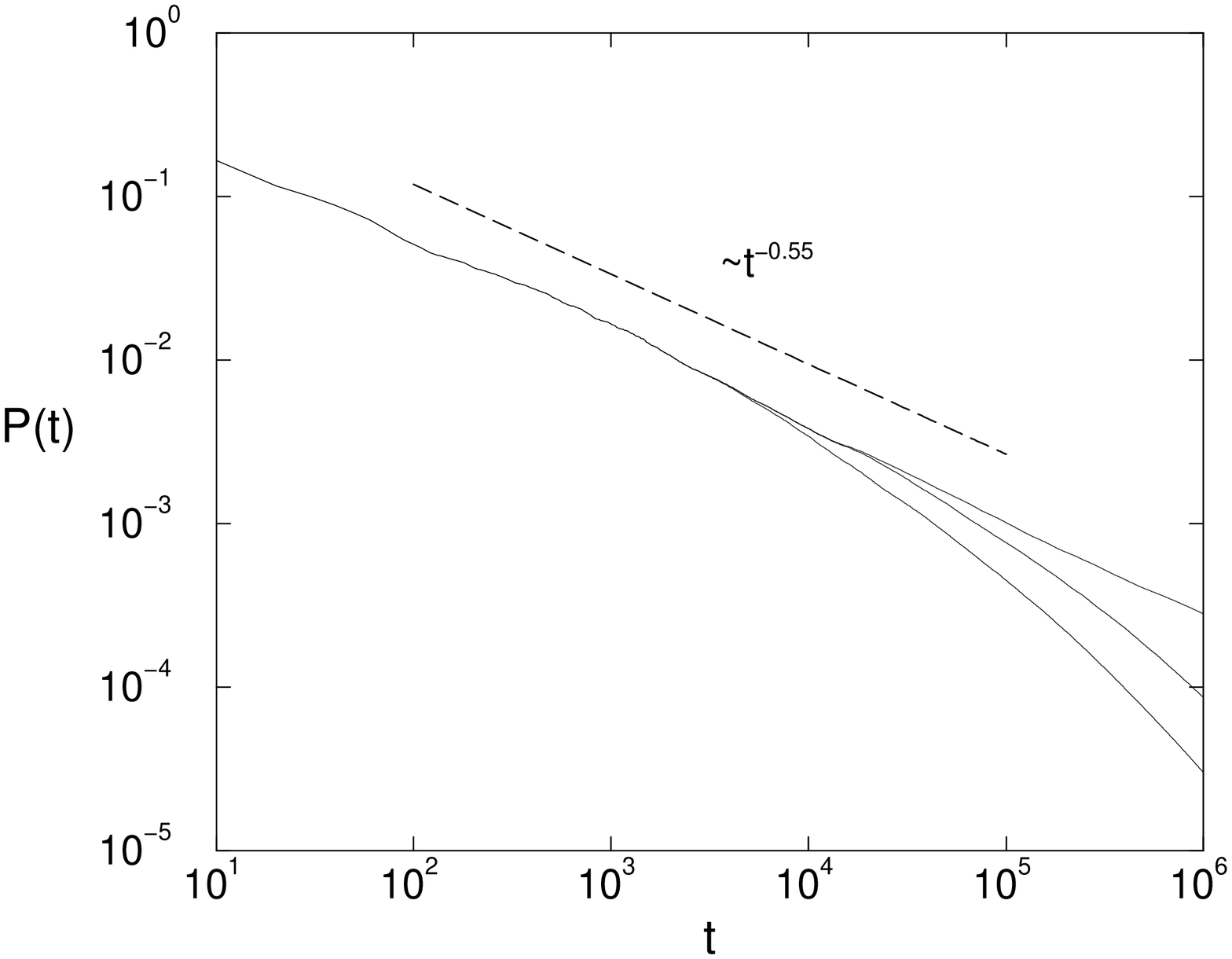,height=9cm,width=6cm,angle=270}
\noindent
{\footnotesize \\
{\bf FIG. 1.}
The survival probability $P(t)$ of the Harper model $(\lambda = 2)$, for
three different sample lengths $L=250, 500, 4000$ exhibits an inverse power-law
$P(t)\sim t^{-\alpha}$. A least squares fit yields $\alpha =0.55\pm
0.05$
in accordance with $D_0^E\simeq 0.5$ and Eq.~(\ref{powlaw}).
}
\end{figure}

The mathematical model we consider is the time-dependent Schr\"odinger
equation
\begin{equation}
\label{eqmo}
i\,{\frac{d\psi_n(t)}{{dt}}}=\,V_n \psi_n(t)+\psi_{n+1}(t)+\psi_{n-1}(t) ,
\end{equation}
where $\psi_n(t)$ is the probability amplitude for an electron to be at site
$n$ of a one-dimensional (1D) sample of length $L$. The on-site potential
$V_n$ is determined by quasi-periodic sequences. We assume absorbing boundary 
conditions at the ends of the sample \cite{note} and initial excitations
in the form of a $\delta-$like packet launched at one of the boundaries, i.e.
$\psi_n(t=0)=\delta_{n,1}$.  Equation~(\ref{eqmo}) has been integrated
numerically using a Cayley scheme \cite{WKG00} with integration time step
$dt=0.1$. We attached 15~additional sites at the ends of the sample
and  erased all components of the wave packet on these sites after
each time step~$dt$. The decay of the norm of the wave packet
obtained in this way was not affected by a further decrease of~$dt$.

\begin{figure}
\hspace*{-0.3cm}
\epsfig{figure=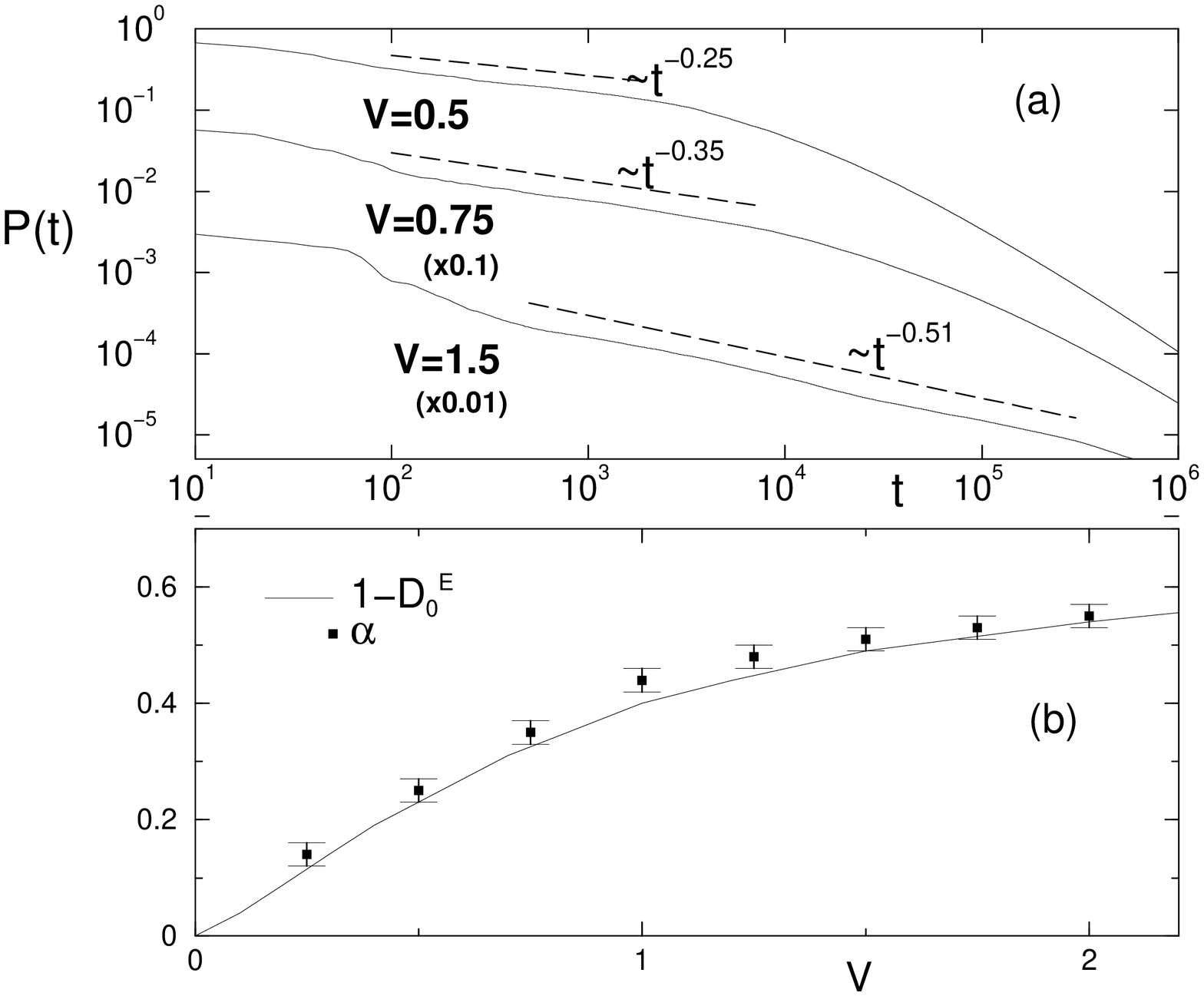,height=8cm,width=8cm,angle=270}
\noindent
{\footnotesize \\
{\bf FIG. 2.}
(a) Survival probabilities $P(t)$ of the Fibonacci model, for three different 
potential strengths $V_1=0.5,~V_2=0.75$ and $V_3=1.5$ showing inverse power-laws 
$P(t)\sim t^{-\alpha}$ (dashed lines).  The sample size is $L=2000$ in all cases.
(b) Power-law exponents $\alpha$ of the survival probability obtained numerically
as a function of the potential strength $V$ for the Fibonacci model. The solid 
line is the theoretical prediction $\alpha = 1-D_0^E$.
}
\end{figure}

We motivate and numerically verify our results using first the well known
Harper model, a paradigm of quasi-periodic 1D systems with a metal-insulator 
transition \cite{GKP95,AA80}. It is described by a tight-binding Hamiltonian 
with an on-site potential given by
\begin{equation}
\label{harper}
V_n=\lambda \cos (2\pi\sigma n ).
\end{equation}
This system effectively describes a particle in a two-dimensional periodic
potential in a uniform magnetic field with $\sigma=a^2eB/hc$ being the number
of flux quanta in a unit cell of area $a^2$. When $\sigma$ is an irrational
number, the period of the effective potential $V_n$ is incommensurate with
the lattice period. The states of the corresponding closed system are extended 
when $\lambda<2$, and the spectrum consists of bands (ballistic regime). For 
$\lambda>2$ the spectrum is point-like and all states are exponentially localized 
(localized regime). The most interesting case corresponds to $\lambda=2$ of 
the metal-insulator transition. At this point, the spectrum is a  Cantor set 
with fractal dimension $D_0^E\leq 0.5$ \cite{frank} while self-similar
fluctuations of the eigenstates exist on all scales \cite{GKP95,AA80}.

\begin{figure}
\hspace*{-0.7cm}
\epsfig{figure=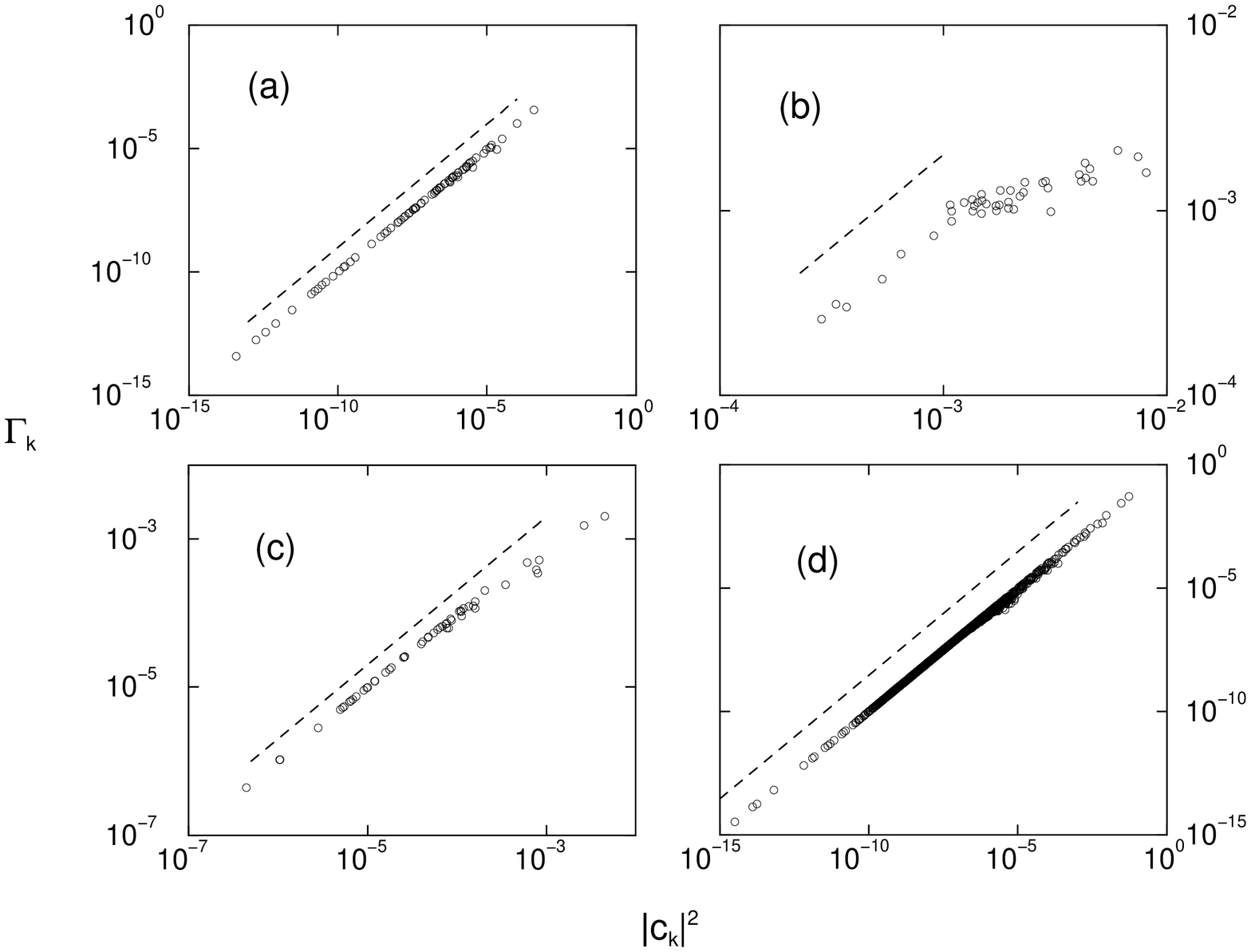,height=9cm,width=7cm,angle=270}
\noindent
{\footnotesize \\
{\bf FIG. 3.}
Resonance widths $\Gamma_k$ as a function of the overlapping elements
$|c_k|^2$. The data are obtained by direct diagonalization of the effective
Hamiltonian ${\cal H}_{eff}$ (see Eq.~(\ref{effec})). The dashed lines
of slope $1$ are shown for comparison demonstrating a linear relation
for small $\Gamma$. In all cases the sample size is $L=1597$, corresponding 
to an approximant of the golden mean $\sigma = \frac {987}{1597}$. (a) Harper 
model for $\lambda=2$; (b) Fibonacci model for $V=0.1$; (c)  Fibonacci model 
for $V=0.5$; and (d)  Fibonacci model for $V=1.5$.
}
\end{figure}

We investigate the survival probability $P(t)$ for the Harper model
at the critical point $\lambda=2$. In our calculations we assume $\sigma$
equal to the golden mean $\sigma_G= ({\sqrt 5}+1)/2$. For this case it is
known that $D_0^E\approx 0.5$ \cite{frank}. The results for various sample
lengths $L$ are shown in Fig.~1. In all cases the survival probability
clearly displays an inverse power law
\begin{equation}
\label{int1a}
P(t)\equiv \sum_{n=1}^L |\psi_n(t)|^2 \sim t^{-\alpha}.
\end{equation}
The best fit to the numerical data yields $\alpha = 0.55\pm 0.05$ in accordance
with Eq.~(\ref{powlaw}).

For a further test of the validity of Eq.~(\ref{powlaw}) we now turn to  the
Fibonacci model, where $D_0^E$ can be varied. Here the potential $V_n$ only 
takes on two values $\pm V$ ($V\neq 0$) that are arranged in a Fibonacci sequence 
\cite{SBGC84}. It was shown that the spectrum is a Cantor set with fractal 
dimension $D_0^E$ depending on $V$ \cite{SBGC84}. In Fig.~2(a) we report some 
of our numerical results for $P(t)$. Again we find a power-law decay $P(t)\sim 
t^{-\alpha}$, where the exponent depends on the potential strength $V$. The 
exponents $\alpha$ extracted for various $V$ are compared with the corresponding 
fractal dimension $D_0^E$ in Fig.~2(b) and confirm the validity of Eq.~(\ref{powlaw}).   

\begin{figure}
\hspace*{-0.7cm}
\epsfig{figure=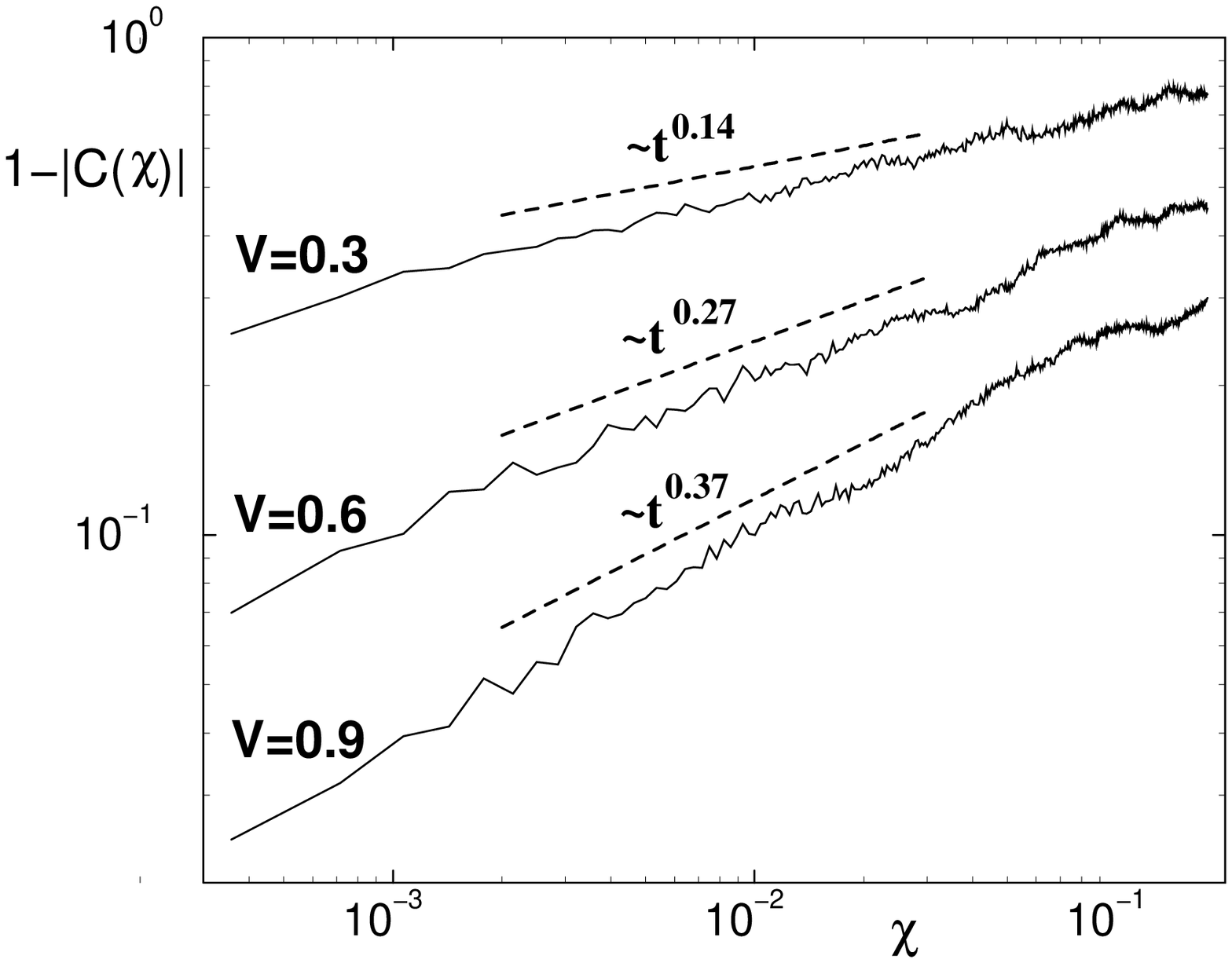,height=5cm,width=9cm,angle=0}
\noindent
{\footnotesize \\
{\bf FIG. 4.}
The scattering autocorrelation function Eq.~(\ref{sscor}) (plotted as
$1-|C(\chi)|$) for some representative $V$ values of the Fibonacci
model compared to the theoretical expectation (dashed line). In
these calculations the sample has length $L=10946$ and is attached to
one lead. The phase of the corresponding scattering matrix $S(E)=
e^{i\Phi(E)}$ was calculated with the help of an iteration relation
developed in Ref.~\cite{OKG00}.
}
\end{figure}

We now want to give a general argument for the validity of Eq.~(\ref{powlaw}). 
The effect of
an open edge for the system described by Eq.~(\ref{eqmo}) can be
simulated by adding the imaginary shift  $i$ to the first
diagonal element of the Hamiltonian matrix \cite{FS97}. Denoting our 
quasi-periodic tight-binding Hamiltonian by $H_L$, this approach
yields an effective Hamiltonian
\begin{equation}
\label{effec}
{\cal H}_{eff} = H_L -i\vec{e}\bigotimes\vec{e},
\end{equation}
where $\vec{e}=(1,0,\ldots ,0)^{~T}$ is an $L-$dimensional vector
that describes at which site  we impose the absorbing boundary
condition. The eigenenergies of the effective Hamiltonian are 
complex ${\cal E}_k = E_k - i\Gamma_k/2$ leading to the decay of the survival
probability $P(t)$. When the on-site potential fulfills $|V_n|>1$ 
the imaginary shift can be considered as a small perturbation of the Hamiltonian
of the closed system. In this case according to the perturbation theory,
$\Gamma_k\sim |\psi_1^k|^2$ holds, where $\psi_n^k$ is an eigenstate
of the closed system with energy $E_k$. The survival probability is then
given by $P(t)\simeq\sum_k |c_k|^2 e^{-\Gamma_k t}$, where
$c_k$ are overlapping elements of the initial state with the
eigenstates $\psi_n^k$. Choosing the initial state to be concentrated at
site $n=1$, we have $|c_k|^2=|\psi_1^k|^2\sim\Gamma_k$. Using
this and converting the sum into an integral we obtain:

\begin{equation}
\label{gamint}
P(t) \sim \int \Gamma\; {\cal P}(\Gamma)\; e^{-\Gamma t}d\Gamma,
\end{equation}
where ${\cal P}(\Gamma)$ is the resonance width distribution. In Ref.~\cite{SOKG00}
it was shown that ${\cal P}(\Gamma) \sim \Gamma^{-(1+D_0^E)}$
for small $\Gamma$. Inserting this expression into Eq.~(\ref{gamint})
one finds the asymptotic power-law decay stated in
Eq.~(\ref{powlaw}). In order to check the validity of the perturbative
arguments we numerically calculate the resonance widths $\Gamma_k$ 
and the overlapping elements $c_k$ for the Harper and Fibonacci models. 
As can be seen from Fig.~3 the prediction $|c_k|^2\sim\Gamma_k$ of the perturbation 
theory holds even for small $V$.

\begin{figure}
\hspace*{-0.7cm}
\epsfig{figure=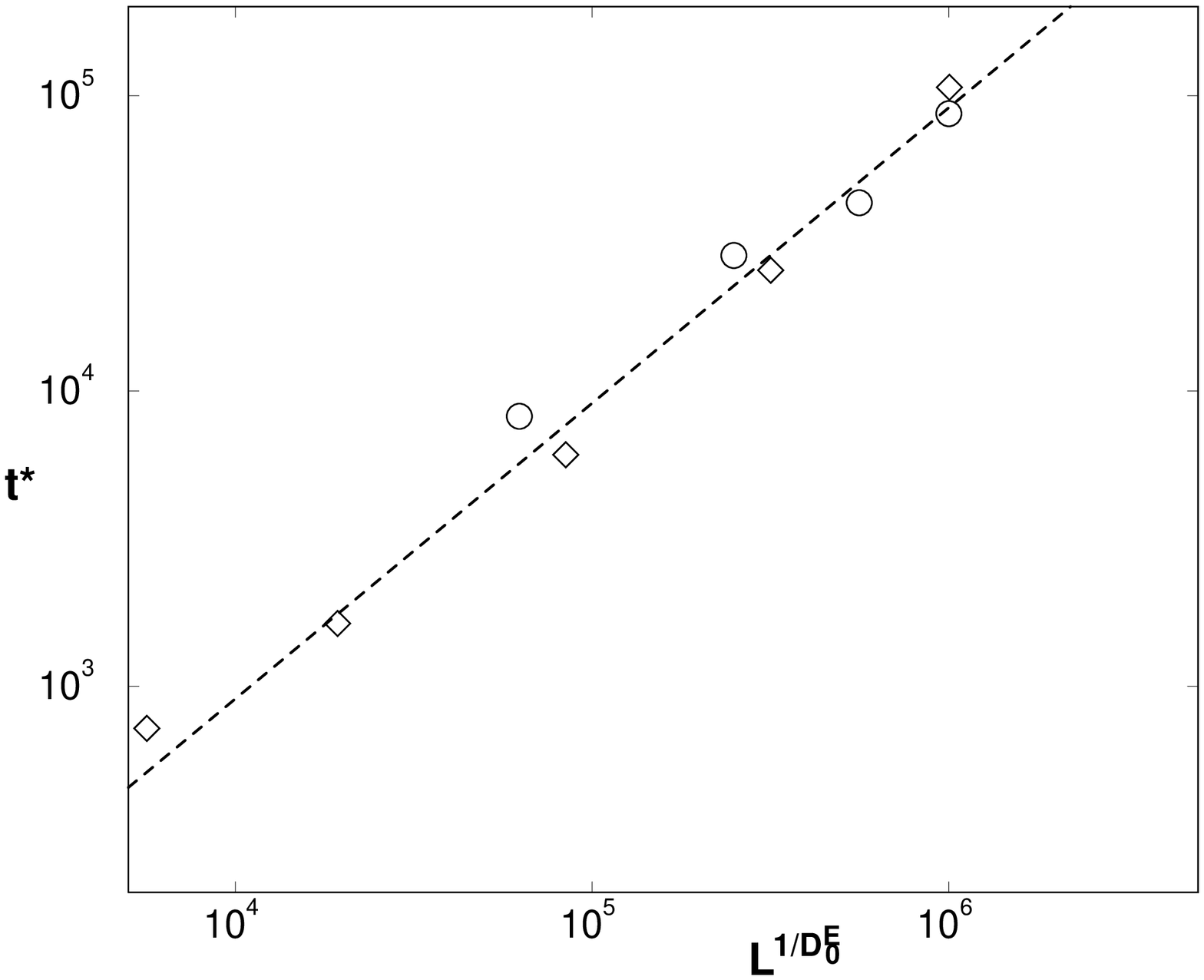,height=9cm,width=5cm,angle=270}
\noindent
{\footnotesize \\
{\bf FIG. 5.}
Dependence of the break-time $t^*$, on the system size $L$ and the box-counting 
dimension $D_0^E$ . ($\circ$) refer to the Harper model at $\lambda=2$ and various 
$L$'s while ($\Diamond$) refer to the Fibonacci model with various $V$'s and $L=2000$. 
The dashed line has slope $1$ and  corresponds to the theoretical expectation Eq.~(\ref{maxt}).
}
\end{figure}

An immediate consequence of Eq.~(\ref{powlaw}) is that the scattering
matrix autocorrelation function $C(\chi)\equiv <S(E)^{\dagger}S(E+\chi)>_E$,
decays in the form of a power law. In particular, using the relation between the
survival probability $P(t)$ and $C(\chi)$ \cite{D00}, we obtain
\begin{equation}
\label{sscor}
1-C(\chi)\sim  \chi \int dt P(t) exp(-i t\chi)\sim \chi^{\alpha},\quad \chi\ll 1.
\end{equation}
Equation ~(\ref{sscor}) is in contrast to the Lorentzian form of $C(\chi)$ 
predicted by RMT for chaotic/ballistic systems \cite{BS88}. Comparison of $C(\chi)$ 
for various $V$-values of the Fibonacci model with the theoretical prediction 
Eq.~(\ref{sscor}) in Fig.~4 shows a nice agreement and provides an
additional check for the validity of Eq.~(\ref{powlaw}).

For finite samples the power-law decay of $P(t)$ (Eq.~(\ref{powlaw})) holds up
to a break time $t^*$, beyond which it turns into an exponential decay. The rate 
of the latter is determined by the smallest resonance width $\Gamma_{min}$ and 
thus $t^*\sim 1/\Gamma_{min}$. An estimation for $t^*$ can be derived as follows.
Imposing the normalization condition for the resonance width distribution and
assuming that the power law ${\cal P}(\Gamma) \sim \Gamma^{-(1+D_0^E)}$
is valid for $\Gamma \ge
 \Gamma_{min}$, i.e $\int_{\Gamma_{min}}^{\infty}d\Gamma {\cal P}(\Gamma) =
L$, we obtain Eq.~(\ref{maxt}). This prediction is verified numerically in
Fig.~5 where we defined $t^*$ as the time where $P(t)$
deviates by 5\% from the power-law decay. We want to point out that
the increase of $t^*$ for decreasing $D_0^E$ is consistent with the
enlargement of the interval where $|c_k|^2\sim\Gamma_k$ holds and
its shift towards smaller values (notice the change of the axes scales in Fig.~3(b)-(d)). 

In summary we find that systems with a fractal spectrum show a power-law 
decay of the survival probability $P(t)\sim t^{-(1-D_0^E)}$. For the finite 
systems of size $L$, this decay can be observed up to a time scale $t^*\sim 
L^{1/D_0^E}$ beyond which an exponential decay sets in. Our predictions should 
be observable in propagation experiments with electrons or classical waves 
in quasi-periodic superlattices or dielectric multilayers \cite{exper}. 
Moreover the connection between $\alpha$, $\gamma$ and $D_0^E$ provides a new 
possibility for experimental measurements of the fractal dimension $D_0^E$ by 
studying current-relaxation processes.

We thank L. Hufnagel, for useful discussions. (T.K) thanks U. Smilansky for
initiating his interest in quantum scattering.


\end{multicols}


\begin{thebibliography}{99}

\vspace*{-1.2cm}

\bibitem{nuclear} C. Mahaux and H. A. Weidenm\"uller, {\it ``Shell Model
Approach in Nuclear Reactions''}, (North-Holland, Amsterdam), (1969);
I. Rotter, Rep. Prog. Phys. {\bf 54}, 635 (1991).


\bibitem{atomic} {\it Atomic Spectra and Collisions in External Fields},
M. H. Nayfeh et. al., eds. (Plenum, New York), Vol. 2 (1989).

\bibitem{molecular} P. Gaspard, in {\it ``Quantum Chaos", Proceedings of
E. Fermi Summer School 1991}, G. Casati et. al., eds. (North-Holland)
307.

\bibitem{mesoscopics} P. W. Brouwer, K. M. Frahm, C. W. Beenakker, Phys.
Rev. Lett.  {\bf 78}, 4737 (1997); M. Titov and C. Beenakker, cond-mat/
0005042 (2000); M. Titov and Yan V. Fyodorov, Phys. Rev. B {\bf 61}, R2444
(2000).

\bibitem{KS00} T. Kottos, U. Smilansky, J. Fortuny, G. Nesti,
Radio Science, {\bf 34} 747, (1999); Tsampikos Kottos and Uzy Smilansky, Phys. 
Rev.  Lett., {\bf 85} 968, (2000).

\bibitem{GSST99}A. Z. Genack, P. Sebbah, M. Stoytchev and B. A. van Tiggelen,
Phys. Rev. Lett., {\bf 82} 715 (1999); J. Stein, H.-J. St\"ockmann, and
U. Stoffregen, Phys. Rev. Lett. {\bf 75}, 53 (1995); Ulrich Kuhl, 
{\it Mikrowellenuntersuchungen an eindimensionalen Streusystemen und in 
zweidimensionalen Kavit\"aten}, Ulrich Kuhl, (Ph.D. Thesis, Philipps-
Universit\"at Marburg, July 1998). 

\bibitem{AKL87} B. L. Altshuler, V. E. Kravtsov, I. V. Lerner, Pisma Zh.
Eksp. Teor. Fiz. {\bf 45}, 160 (1987) [ JETP Lett. {\bf 45}, 199 (1987)];
B. L. Altshuler, V. N. Prigodin, Zh. Eksp. Teor. Fiz. {\bf 95}, 348 (1989) 
[ Sov. Phys. JETP {\bf 68}, 409 (1980)]

\bibitem{MK95} B. A. Muzykantskii and D. E. Khmelnitskii, Phys. Rev. B
{\bf 51}, 5480 (1995).

\bibitem{M95} A. D. Mirlin, Phys. Rep. {\bf 326}, 259 (2000).

\bibitem{FE95} V. I. Fal'ko and K. B. Efetov, Phys. Rev. B
{\bf 52}, 17413 (1995); V. I. Fal'ko and K. B. Efetov, Europhys. Lett. 
{\bf 32}, 627 (1995)
\bibitem{SA97} I. E. Smolyarenko and B. L. Altshuler, Phys. Rev. B {\bf 55},
10451 (1997).

\bibitem{SS97} D. V. Savin and V. V. Sokolov, Phys. Rev. E {\bf 56}, R4911
(1997); M. Gl\"uck, A. R. Kolovsky and H. J. Korsch, nlin.CD/0005065.

\bibitem{D00}F. M. Dittes, Phys. Reports {\bf 339}, 215 (2000).

\bibitem{CMS99} G. Casati, G. Maspero, and D. L. Shepelyansky, Phys. Rev Lett.
{\bf 82}, 524 (1999); ibid, {\bf 84}, 4088 (2000).

\bibitem{HKW00} G. Casati, I. Guarneri, and G. Maspero, Phys. Rev Lett.
{\bf 84}, 63 (2000); L. Hufnagel, R. Ketzmerick, M. Weiss, cond-mat/0009010.

\bibitem{GKP95} T. Geisel, R. Ketzmerick and G. Petschel, Phys.  Rev. Lett.
{\bf 66}, 1651 (1991); T. Geisel, R. Ketzmerick and G. Petschel, in {\it Quantum
Chaos; Between Order and Disorder}, J. Casati and B. Chirikov, eds. (Cambridge)
634-659.

\bibitem{P65}C. E. Porter, {\it Statistical Theory of Spectral Fluctuations},
Academic Press, New York (1965); B. L. Altshuler and B. I. Shklovskii, Zh. 
Eksp. Teor. Fiz. {\bf 91}, 220 (1986) [Sov. Phys. JETP {\bf 64}, 127 (1986)].

\bibitem{AA80} P. G. Harper, Proc. Roy. Soc. Lond. {\bf A68}, 874 (1955); S.
Aubry, G. Andre, Ann. Israel Phys. Soc. {\bf 3}, 133 (1980); S. N. Evangelou, 
J.-L. Pichard, Phys. Rev. Lett. {\bf 84}, 1643 (2000); 
F. Pi\'echon, Phys. Rev. Lett., {\bf 76} 4372,
(1996).

\bibitem{SBGC84}D. Shechtman, I. Blech, D. Gratias, and J. V. Cahn, Phys.
Rev. Lett. {\bf 53}, 1951 (1984); M. Kohmoto, L. P. Kadanoff and G. Tang, Phys.
Rev. Lett. {\bf 50} 1870 (1983); J. M. Luck and D. Petritis, J. Stat. Phys.
{\bf 42}, 289 (1986); J. Bellisard, B. Iochum, E. Scoppola, and D. Testard,
Comm. Math. Phys. {\bf 125}, 527 (1989).

\bibitem{GKP91} P. Leboeuf, J. Kurchan, M. Feingold, and D. P. Arovas, Phys.
Rev. Lett. {\bf 65}, 3076 (1990); T. Geisel, R. Ketzmerick and G. Petschel,
Phys.  Rev. Lett. {\bf 67}, 3635 (1991); R. Artuso, F. Borgonovi, I. Guarneri,
L. Rebuzzini, and G. Casati, Phys. Rev. Lett. {\bf 69}, 3302 (1992);

\bibitem{note} Our results valid also for the case of a semi-infinite lattice
with one absorbing boundary. 

\bibitem{WKG00}M. Weiss, Tsampikos Kottos, and T. Geisel, cond-mat/0005339;
W. H. Press and S. A. Teukolsky et al., {\it
``Numerical Recipes in Fortran 90''}, (Cambridge University Press), (1996);

\bibitem{frank} Y.~Last,~Commun.~Math.~Phys.~{\bf 164},~421~(1994);
R. Ketzmerick, K. Kruse, F. Steinbach, and T. Geisel,Phys. Rev. B {\bf 58},
9881, (1998).

\bibitem{FS97} Y. V. Fyodorov, H-J Sommers, J. Math. Phys. {\bf 38}, 1918
(1997); M. Terraneo, and I. Guarneri, Eur. Phys. J. B {\bf 18}, 303 (2000). 

\bibitem{SOKG00} F. Steinbach, A. Ossipov, Tsampikos Kottos, and T. Geisel,
Phys. Rev. Lett. {\bf 85}, 4426 (2000).

\bibitem{OKG00} A. Ossipov, Tsampikos Kottos, and Theo Geisel, Phys. Rev. B,
{\bf 61} 11411, (2000).

\bibitem{BS88} R. Bl\"umel and U. Smilansky, Phys. Rev. Lett., {\bf 60} 477
(1988).

\bibitem{exper} U. Kuhl and H.-J. St\"ockmann, Phys. Rev. Lett. {\bf 80}, 
3232 (1998).

\end{thebibliography}
\end{document}